\documentclass[12pt]{article}

\usepackage{amssymb}
\usepackage{amsfonts}
\usepackage{amsmath}

\begin{document}

\title{On Shor's channel extension and constrained channels}
\author{A. S. Holevo\thanks{
Steklov Mathematical Institute, 119991 Moscow, Russia}, M.E.Shirokov\thanks{
Moscow Institute of Physics and Technology, 141700 Moscow, Russia}}
\date{}
\maketitle

\begin{abstract}
Several equivalent formulations of the additivity conjecture for constrained
channels, which formally is substantially stronger than the unconstrained
additivity, are given. To this end a characteristic property of the optimal
ensemble for such a channel is derived, generalizing the maximal distance
property. It is shown that the additivity conjecture for constrained
channels holds true for certain nontrivial classes of channels. After giving
an algebraic formulation for the Shor's channel extension, its main
asymptotic property is proved. It is then used to show that additivity for
two constrained channels can be reduced to the same problem for
unconstrained channels, and hence, ``global'' additivity for channels with
arbitrary constraints is equivalent to additivity without constraints.

\textit{Running title:} Shor's channel extension and constrained channels
\end{abstract}

\section{Introduction}

In the recent paper \cite{Sh-e-a-q} Shor gave arguments which show that
conjectured additivity properties for several quantum information
quantities, such as the minimal output entropy, the Holevo capacity (in what
follows $\chi $-capacity) and the entanglement of formation are in fact
equivalent. An important new tool in these arguments is the construction of
special extension $\widehat{\Phi }$ for an arbitrary channel $\Phi $ which
has desired properties lacking for the initial channel. In this paper we
show that this extension allows us to deal with the additivity conjecture
for quantum channels with constrained inputs. Introducing input constraints
provides greater flexibility in the treatment of the additivity conjecture.
In a sense, Shor's channel extension plays a role of the Lagrange function
in optimization for the additivity questions. On the other hand, while \cite
{Sh-e-a-q} deals with the ``global'' additivity, i.e. properties valid for
all possible channels, in this paper we make emphasis on results valid for
individual channels.

We start with giving several equivalent formulations of the additivity
conjecture for constrained channels (theorem 1), which formally is
substantially stronger than the unconstrained additivity. To this end a
characteristic property of the optimal ensemble for such a channel is
derived (proposition 1), generalizing the maximal distance property \cite
{Sch-West-1}. It is shown that the additivity conjecture for constrained
channels holds true for certain nontrivial classes of channels (proposition
2). After giving an algebraic formulation for the Shor's channel extension
\cite{Sh-e-a-q}, its main property (proposition 3) is proved. It is then
used to show that additivity for two constrained channels can be reduced to
the same problem for unconstrained channels, and hence, global additivity
for channels with arbitrary constraints is equivalent to global additivity
without constraints (theorem 2 and corollaries). Further results in this
direction can be found in \cite{H-Sh}.

\section{Basic quantities}

Let $\mathcal{H},\mathcal{H}^{\prime }$ be finite dimensional Hilbert spaces
and let $\Phi :\mathfrak{S}(\mathcal{H})\mapsto \mathfrak{S}(\mathcal{H}
^{\prime })$ be a channel, where $\mathfrak{S}(\mathcal{H})$ \ denotes the
set of states (density operators) in $\mathcal{H}$. Let $\{\pi _{i}\}$ be a
finite probability distribution and $\{\rho _{i}\}$ a collection of states
in $\mathfrak{S}(\mathcal{H}),$ then the collection $\{\pi _{i},\rho _{i}\}$
is called \textit{ensemble, }and $\rho _{\mathrm{av}}=\sum_{i}\pi _{i}\rho
_{i}$ is its \textit{average}.

An important entropic characteristic of ensemble is defined by
\begin{equation}
\chi _\Phi \left( \{\pi _i,\rho _i\}\right) =H\left( \sum_i\pi _i\Phi \left(
\rho _i\right) \right) -\sum_i\pi _iH\left( \Phi \left( \rho _i\right)
\right) ,  \label{chiq}
\end{equation}
where $H\left( \cdot \right) $ is the von Neumann entropy. Following \cite
{MSW}, we denote
\begin{equation*}
\chi _\Phi (\rho )=\max_{\rho _{\mathrm{av}}=\rho }\chi _\Phi (\{\pi _i,\rho
_i\}).\vspace{10pt}
\end{equation*}
Notice that
\begin{equation}
\chi _\Phi (\rho )=H\left( \Phi \left( \rho \right) \right) -\hat{H}_\Phi
\left( \rho \right) ,  \label{chihat}
\end{equation}
where
\begin{equation*}
\hat{H}_\Phi \left( \rho \right) =\min_{\rho _{\mathrm{av}}=\rho }\sum_i\pi
_iH\left( \Phi \left( \rho _i\right) \right) .
\end{equation*}
The function\ $\hat{H}_\Phi \left( \rho \right) $ is the \textit{convex
closure} \cite{JT}, \cite{AB} (or the convex roof, cf. \cite{uhl}) of the
\textit{output entropy} $H\left( \Phi \left( \rho \right) \right) ,$ which
is continuous concave function. The function\ $\hat{H}_\Phi \left( \rho
\right) $ is a natural generalization of the entanglement of formation and
coincides with it when the channel $\Phi $ is a partial trace. The
continuity of $\hat{H}_\Phi \left( \rho \right) $ follows from the MSW
correspondence \cite{MSW} and the continuity of the entanglement of
formation \cite{N}. Thus the function $\chi _\Phi (\rho )$ (briefly $\chi $
-function) is itself continuous and concave on $\mathfrak{S}(\mathcal{H})$.

Consider the constraint on the ensemble $\{\pi _{i},\rho _{i}\}$ defined by
the requirement $\rho _{\mathrm{av}}\in \mathcal{A}$, where $\mathcal{A}$ is
a closed subset of $\mathfrak{S}(\mathcal{H}).$ A particular case is linear
constraint, where the subset $\mathcal{A}^{l}$ is defined by the inequality $
\mathrm{Tr}A\rho _{\mathrm{av}}\leq \alpha $ for some positive operator $A$
and a number $\alpha \geq 0$. Define the $\chi $-\textit{capacity} of the $
\mathcal{A}$-constrained channel $\Phi $ by
\begin{equation}
\bar{C}(\Phi ;\mathcal{A})=\max_{\rho \in \mathcal{A}}\chi _{\Phi }(\rho
)=\max_{\rho _{\mathrm{av}}\in \mathcal{A}}\chi _{\Phi }(\{\pi _{i},\rho
_{i}\}).  \label{ccap}
\end{equation}
In case of the linear constraint $\mathcal{A}^{l}$ we also use the notation $
\bar{C}(\Phi ;A,\alpha ).$ Note that the $\chi $-capacity for the
unconstrained channel is $\bar{C}(\Phi )=\bar{C}(\Phi ;\mathfrak{S }(
\mathcal{H}))$.

\textbf{Lemma 1.} \textit{For arbitrary channel} $\Phi :\mathfrak{S}(
\mathcal{H})\mapsto \mathfrak{S}(\mathcal{H}^{\prime })$ \textit{and
arbitrary density operator }$\rho _{0}$\textit{\ of full rank there exists a
positive operator }$A\leq I_{\mathcal{H}}$ in $\mathfrak{B}(\mathcal{H})$
\textit{such that} $\rho _{0}$ \textit{is the maximum point of the function }
$\chi_{\Phi }(\rho )$\textit{\ under the condition }$\mathrm{Tr}A\rho \leq
\alpha $\textit{, where }$\alpha =\mathrm{Tr}A\rho _{0}$.

The statement of the lemma is intuitively clear, but its proof (see
Appendix, I) requires an argument from the convex analysis due to the fact
that the function $\chi_{\Phi } (\rho )$ may not be smooth.

\section{Optimal ensembles}

An ensemble $\{\pi _{i},\rho _{i}\}$ on which the maximum in (\ref{ccap}) is
achieved is called an \textit{optimal ensemble} for the $\mathcal{A}$
-constrained channel $\Phi $. The following proposition generalizes the
\textit{maximal distance property }of optimal ensembles for unconstrained
channels \cite{Sch-West-1}.

\textbf{Proposition 1.} \textit{Let }$\mathcal{A}$\textit{\ be a closed
convex set. The ensemble }$\{\pi _{i},\rho _{i}\}$\textit{\ with the average
state }$\rho _{\mathrm{av}}\in \mathcal{A}$\textit{\ is optimal for the }$
\mathcal{A}$\textit{-constrained channel }$\Phi $\textit{\ if and only if }
\begin{equation*}
\sum_{j}\mu _{j}H(\Phi (\omega _{j})\Vert \Phi (\rho _{\mathrm{av}}))\leq
\chi _{\Phi }(\{\pi _{i},\rho _{i}\})
\end{equation*}
\textit{for any ensemble }$\{\mu _{j},\omega _{j}\}$\textit{\ with the
average }$\omega _{\mathrm{av}}\in \mathcal{A}$, \textit{where} $H(\cdot
\Vert \cdot ) $\textit{\ is the relative entropy.}

\textit{Proof.} The proof generalizes the argument in \cite{Sch-West-1} by
considering variations of the initial ensemble involving not a single
component but the whole ensemble.

Let $\{\pi _i,\rho _i\}_{i=1}^n$ and $\{\mu _j,\omega _j\}_{j=1}^m$ be two
ensembles with the averages $\rho _{\mathrm{av}}$ and $\omega _{\mathrm{av}}$
contained in $\mathcal{A}$. Consider the variation of the first ensemble by
mixing it with the second one with the weight coefficient $\eta $. The
modified ensemble
\begin{equation*}
\Sigma ^\eta =\{(1-\eta )\pi _1\rho _1,...,(1-\eta )\pi _n\rho _n,\eta \mu
_1\omega _1,...,\eta \mu _m\omega _m\}
\end{equation*}
has the average $\rho _{\mathrm{av}}^\eta =(1-\eta )\rho _{\mathrm{av}}+\eta
\omega _{\mathrm{av}}\in \mathcal{A}$ (by convexity). Using the relative
entropy expression for the quantity (\ref{chiq}), we have
\begin{equation}
\chi _\Phi \left( \Sigma ^\eta \right) =(1-\eta )\sum_{i=1}^n\pi _iH(\Phi
(\rho _i)\Vert \Phi (\rho _{\mathrm{av}}^\eta ))+\eta \sum_{j=1}^m\mu
_jH(\Phi (\omega _j)\Vert \Phi (\rho _{\mathrm{av}}^\eta )).  \label{m-chi}
\end{equation}
Applying Donald's identity \cite{Sch-West-1}, \cite{Sch-West-2} to the
original ensemble we obtain
\begin{equation*}
\sum_{i=1}^n\pi _iH(\Phi (\rho _i)\Vert \Phi (\rho _{\mathrm{av}}^\eta
))=\chi _\Phi (\Sigma ^0)+H(\Phi (\rho _{\mathrm{av}})\Vert \Phi (\rho _{
\mathrm{av}}^\eta )).
\end{equation*}
Substitution of the above expression into (\ref{m-chi}) gives
\begin{equation}
\begin{array}{c}
\chi _\Phi \left( \Sigma ^\eta \right) =\chi _\Phi (\Sigma ^0)+(1-\eta
)H(\Phi (\rho _{\mathrm{av}})\Vert \Phi (\rho _{\mathrm{av}}^\eta )) \\
\\
+\eta \left[ \sum\limits_{j=1}^m\mu _jH(\Phi (\omega _j)\Vert \Phi (\rho _{
\mathrm{av}}^\eta ))-\chi _\Phi (\Sigma ^0)\right] .\label{m-chi-1}
\end{array}
\end{equation}
Applying Donald's identity to the modified ensemble we obtain
\begin{equation*}
\begin{array}{c}
(1-\eta )\sum\limits_{i=1}^n\pi _iH(\Phi (\rho _i)\Vert \Phi (\rho _{\mathrm{
\ av}}))+\eta \sum\limits_{j=1}^m\mu _jH(\Phi (\omega _j)\Vert \Phi (\rho _{
\mathrm{av}})) \\
\\
=\chi _\Phi \left( \Sigma ^\eta \right) +H(\Phi (\rho _{\mathrm{av}}^\eta
)\Vert \Phi (\rho _{\mathrm{av}}))
\end{array}
\end{equation*}
and hence
\begin{equation}
\begin{array}{c}
\chi _\Phi \left( \Sigma ^\eta \right) =\chi _\Phi \left( \Sigma ^0\right)
-H(\Phi (\rho _{\mathrm{av}}^\eta )\Vert \Phi (\rho _{\mathrm{av}})) \\
\\
+\eta \left[ \sum\limits_{j=1}^m\mu _jH(\Phi (\omega _j)\Vert \Phi (\rho _{
\mathrm{av}}))-\chi _\Phi \left( \Sigma ^0\right) \right] .\label{m-chi-2}
\end{array}
\end{equation}
Since the relative entropy is nonnegative, the expressions (\ref{m-chi-1})
and (\ref{m-chi-2}) imply the following inequalities for the quantity $
\Delta \chi _\Phi =\chi _\Phi $ $\left( \Sigma ^\eta \right) -\chi _\Phi $ $
\left( \Sigma ^0\right) $:
\begin{equation}
\begin{array}{c}
\eta \left[ \sum\limits_{j=1}^m\mu _jH(\Phi (\omega _j)\Vert \Phi (\rho _{
\mathrm{av}}^\eta ))-\chi _\Phi \left( \Sigma ^0\right) \right] \\
\leq \Delta \chi _\Phi \leq \\
\eta \left[ \sum\limits_{j=1}^m\mu _jH(\Phi (\omega _j)\Vert \Phi (\rho _{
\mathrm{av}}))-\chi _\Phi \left( \Sigma ^0\right) \right] .
\end{array}
\label{d-chi}
\end{equation}
Now the proof of the proposition is straightforward. If
\begin{equation*}
\sum_j\mu _jH(\Phi (\omega _j)\Vert \Phi (\rho _{\mathrm{av}}))\leq \chi
_\Phi \left( \Sigma ^0\right)
\end{equation*}
for any ensemble $\{\mu _j,\omega _j\}$ of states in $\mathfrak{S}(\mathcal{
\ \ \ H })$ with the average $\omega _{\mathrm{av}}\in \mathcal{A}$, then by
the second inequality in (\ref{d-chi}) with $\eta =1$ we have
\begin{equation*}
\chi _\Phi (\{\mu _j,\omega _j\})=\chi _\Phi (\Sigma ^1)\leq \chi _\Phi
\left( \Sigma ^0\right) =\chi _\Phi (\{\pi _i,\rho _i\}),
\end{equation*}
which means optimality of the ensemble $\{\pi _i,\rho _i\}$.

To prove the converse, suppose $\{\pi _{i},\rho _{i}\}$ is an optimal
ensemble and there exists an ensemble $\{\mu _{j},\omega _{j}\}$ such that
\begin{equation*}
\sum_{j}\mu _{j}H(\Phi (\omega _{j})\Vert \Phi (\rho _{\mathrm{av}}))>\chi
_{\Phi }\left( \Sigma ^{0}\right) .
\end{equation*}
By continuity of the relative entropy, there is $\eta >0$ such that
\begin{equation*}
\sum_{j}\mu _{j}H(\Phi (\omega _{j})\Vert \Phi (\rho _{\mathrm{av}}^{\eta
}))>\chi _{\Phi }\left( \Sigma ^{0}\right) .
\end{equation*}
By the first inequality in (\ref{d-chi}), this means that $\chi _{\Phi }$ $
\left( \Sigma ^{\eta }\right) >\chi _{\Phi }$ $\left( \Sigma ^{0}\right) $
in contradiction with the optimality of the ensemble $\{\pi _{i},\rho _{i}\}$
. $\square $

\textbf{Corollary 1.} \textit{Let } $\rho _{\mathrm{av}}$\textit{\ be the
average of an optimal ensemble for the }$\mathcal{A}$\textit{-constrained
channel }$\Phi ,$\textit{\ then}
\begin{equation*}
\bar{C}(\Phi ;\mathcal{A})=\chi _{\Phi }(\rho _{\mathrm{av}})\geq \chi
_{\Phi }(\rho )+H(\Phi (\rho )\Vert \Phi (\rho _{\mathrm{av}})),\quad
\forall \rho \in \mathcal{A}.
\end{equation*}
\textit{Proof.} Let $\{\pi _{i},\rho _{i}\}$ be an arbitrary ensemble such
that $\sum_{i}\pi _{i}\rho _{i}=\rho \in \mathcal{A}$. By proposition 1
\begin{equation*}
\sum_{i}\pi _{i}H(\Phi (\rho _{i})\Vert \Phi (\rho _{\mathrm{av}}))\leq \chi
_{\Phi }(\rho _{\mathrm{av}}).
\end{equation*}
This inequality and Donald's identity
\begin{equation*}
\sum_{i}\pi _{i}H(\Phi (\rho _{i})\Vert \Phi (\rho _{\mathrm{av}}))=\chi
_{\Phi }(\{\pi _{i},\rho _{i}\})+H(\Phi (\rho )\Vert \Phi (\rho _{\mathrm{av}
})).
\end{equation*}
complete the proof. $\square $

\section{Additivity for constrained channels}

Let $\Psi :\mathfrak{S}(\mathcal{K})\mapsto \mathfrak{S}(\mathcal{K}^{\prime
})$ be another channel with the constraint, defined by a closed subset $
\mathcal{B}\subset \mathfrak{S}(\mathcal{K})$. For the channel $\Phi\otimes
\Psi $ we consider the constraint defined by the requirements $\sigma _{
\mathrm{av}}^{\Phi }:=\mathrm{Tr}_{\mathcal{K}}\sigma _{\mathrm{av}}\in
\mathcal{A}$ and $\sigma _{\mathrm{av}}^{\Psi }:=\mathrm{Tr}_{\mathcal{H}
}\sigma_{\mathrm{av}}\in \mathcal{B}$, where $\sigma _{\mathrm{av}}$ is the
average state of an input ensemble $\{\mu _{i},\sigma _{i}\}$. The closed
subset of $\mathfrak{S}(\mathcal{H}\otimes \mathcal{K})$ defined by the
above requirements will be denoted $\mathcal{A}\otimes \mathcal{B}$.

We conjecture the following additivity property for constrained channels
\begin{equation}
\bar{C}\left( \Phi \otimes \Psi ;\mathcal{A}\otimes \mathcal{B}\right) =\bar{
C}(\Phi ;\mathcal{A})+\bar{C}(\Psi ;\mathcal{B}).  \label{addit}
\end{equation}
The usual additivity conjecture for unconstrained channels is obtained by
setting $\mathcal{A=}\mathfrak{S}(\mathcal{H}),$ $\mathcal{B=}\mathfrak{S}(
\mathcal{K}).$

\textbf{Theorem 1}\textit{. Let }$\Phi $\textit{\ and }$\Psi $\textit{\ be
fixed channels. The following properties are equivalent:}

$\mathit{(i)}$\textit{\ \ equality (\ref{addit}) holds for arbitrary closed }
$\mathcal{A}$\textit{\ and }$\mathcal{B}$\textit{;}

$\mathit{(ii)}$\textit{\ \ equality (\ref{addit}) holds for arbitrary linear
constraints} $\mathcal{A}^{l}$\textit{\ and }$\mathcal{B}^{l}$ \textit{;}

$\mathit{(iii)}$\textit{\ for arbitrary }$\sigma \in \mathfrak{S}(\mathcal{H}
\otimes \mathcal{K})$
\begin{equation}
\chi _{\Phi \otimes \Psi }(\sigma )\leq \chi _{\Phi }(\sigma ^{\Phi })+\chi
_{\Psi }(\sigma ^{\Psi });  \label{sub-add}
\end{equation}

$\mathit{(iv)}$\textit{\ for arbitrary }$\sigma \in \mathfrak{S}(\mathcal{H}
\otimes \mathcal{K})$
\begin{equation}
\hat{H}_{\Phi \otimes \Psi }(\sigma )\geq \hat{H}_\Phi (\sigma ^\Phi )+ \hat{
H}_\Psi (\sigma ^\Psi );  \label{hat}
\end{equation}

These are also equivalent to the corresponding additivity properties of $
\chi _\Phi $ and $\hat{H}_\Phi $ for tensor product states. By using the MSW
correspondence the case of $\hat{H}_\Phi $ can be reduced to entanglement of
formation, for which this was established in \cite{Sh-e-a-q}, \cite{P}.

\textit{Proof.} $(i)\Rightarrow (ii)$ is obvious. $(ii)\Rightarrow (i)$ can
be proved by double application of the following lemma.

\textbf{Lemma 2.} \textit{The equality (\ref{addit}) holds for fixed closed}
$\mathcal{B}$ \textit{and arbitrary closed} $\mathcal{A}$\textit{\ if it
holds for the set }$\mathcal{B}$ \textit{and arbitrary linear constraint} $
\mathcal{A}^{\mathit{l}}$, \textit{defined by the inequality} $\mathrm{Tr}
A\rho \leq \alpha $ \textit{with a positive operator }$A$\textit{\ and a
number }$\alpha $\textit{\ such that there exists a state }$\rho ^{\prime }$
\textit{\ with} $\mathrm{Tr}A\rho ^{\prime }<\alpha $.

\textit{Proof.} Assume that the equality (\ref{addit}) holds for the set $
\mathcal{B}$ and arbitrary set $\mathcal{A}^{\mathit{l}}$, satisfying the
above condition. It is sufficient to prove that
\begin{equation}
\chi _{\Phi \otimes \Psi }(\sigma )\leq \chi _{\Phi }(\sigma ^{\Phi })+\bar{
C }(\Psi ;\mathcal{B})  \label{sp-add-ineq}
\end{equation}
for any $\sigma \in \mathfrak{S}(\mathcal{H}\otimes \mathcal{K})$ such that $
\sigma ^{\Psi }\in \mathcal{B}$. \ Due to continuity of the $\chi $
-function, it is sufficient to prove (\ref{sp-add-ineq}) for a state $\sigma
$ with partial trace $\sigma ^{\Phi }$ of full rank. For the state $\sigma
^{\Phi }$ we can choose a positive operator $A$ in $\mathfrak{B}(\mathcal{H}
) $ in accordance with lemma 1. Let $\mathcal{A}^{\mathit{l}}=\{\rho \in
\mathfrak{S}(\mathcal{H})\,|\,\mathrm{Tr}A\rho \leq \alpha =\mathrm{Tr}
A\sigma ^{\Phi }\}$. The full rank of $\sigma ^{\Phi }$ guarantees the
existence of a state $\rho ^{\prime }$ such that $\mathrm{Tr}A\rho ^{\prime
}<\alpha =\mathrm{Tr}A\sigma ^{\Phi }$. Let $\omega $ be the average state
of the optimal ensemble for the $\mathcal{B}$-constrained channel $\Psi $.
Due to the above assumption the state $\sigma ^{\Phi }\otimes \omega $ is
the average state of the optimal ensemble for $\mathcal{A}^{\mathit{l}
}\otimes \mathcal{B}$-constrained channel $\Phi \otimes \Psi $. But it is
clear that this ensemble will also be optimal for $\{\sigma ^{\Phi
}\}\otimes \mathcal{B}$-constrained channel $\Phi \otimes \Psi $ and, hence,
(\ref{sp-add-ineq}) is true. $\square $

$(i)\Rightarrow (iv).$ Fix the states $\rho$ and $\omega$ and take $\mathcal{
\ A}=\{\rho \}$, $\mathcal{B}=\{\omega \},$ then (\ref{addit}) becomes
\begin{equation}
\bar{C}\left( \Phi \otimes \Psi ;\{\rho \}\otimes \{\omega \}\right) =\bar{
C }(\Phi ;\{\rho \})+\bar{C}(\Psi ;\{\omega \}).  \label{ind}
\end{equation}
This implies existence of unentangled ensemble with the average $\rho
\otimes \omega $, which is optimal for the $\{\rho \}\otimes \{\omega \}$
-constrained channel $\Phi \otimes \Psi $. By corollary 1 we have
\begin{equation}
\!\!\!\chi _{\Phi \otimes \Psi }(\rho\otimes\omega)\!=\!\chi _{\Phi }(\rho
)+\chi _{\Psi}(\omega )\!\geq\!\chi _{\Phi\otimes \Psi}(\sigma )+H((\Phi
\otimes \Psi )(\sigma )\Vert \Phi (\rho )\otimes \Psi (\omega))\!
\label{opt-c}
\end{equation}
for any state $\sigma \in \mathfrak{S}(\mathcal{H})\otimes \mathfrak{S}(
\mathcal{K})$ such that $\sigma ^{\Phi }=\rho $ and $\sigma ^{\Psi }=\omega $
. Note that
\begin{equation}
\!\!\!H((\Phi \otimes \Psi )(\sigma )\Vert \Phi (\rho )\otimes \Psi (\omega
))=H(\Phi (\rho ))+H(\Psi (\omega ))-H((\Phi \otimes \Psi )(\sigma )).
\label{s-c-r-e-exp}
\end{equation}
The inequality (\ref{opt-c}) together with (\ref{s-c-r-e-exp}) and (\ref
{chihat}) implies (\ref{hat}).

$(iv)\Rightarrow (iii)$ obviously follows from the definition of the $\chi $
-function and subadditivity of the (output) entropy.

$(iii)\Rightarrow (i)$. From the definition of the $\chi $-capacity and (\ref
{sub-add})

\begin{equation*}
\bar{C}\left( \Phi \otimes \Psi ;\mathcal{A}\otimes \mathcal{B}\right) \leq
\bar{C}(\Phi ;\mathcal{A})+\bar{C}(\Psi ;\mathcal{B}).
\end{equation*}
Since the converse inequality is obvious, there is equality here. $\square $

\textbf{Remark 1.} The additivity of the $\chi -$capacity for arbitrarily
constrained channels is formally substantially stronger than the usual
unconstrained additivity. Indeed, the latter holds trivially for channels
that are (unconstrained) partial traces, but the additivity for constrained
partial traces, by the MSW correspondence, would imply validity of the
global additivity conjecture.

The following proposition implies that the set of quantum channels
satisfying the properties in theorem 1 is nonempty. We shall use the
following obvious statement

\textbf{Lemma 3.}\textit{\ Let }$\{\Phi _{j}\}_{j=1}^{n}$\textit{\ be a
collection of channels from }$\mathfrak{S}(\mathcal{H})$\textit{\ into }$
\mathfrak{S}(\mathcal{H}_{j})$, and let $\{q_{j}\}_{j=1}^{n}$\textit{\ be a
probability distribution. Then for the channel }$\Phi
=\bigoplus_{j=1}^{n}q_{j}\Phi _{j}$ \textit{\ from }$\mathfrak{S}(\mathcal{H}
)$\textit{\ into }$\mathfrak{S}(\bigoplus_{j=1}^{n}\mathcal{H}_{j})$\textit{
\ \ one has}
\begin{equation*}
\chi _{\Phi }\left( \{\rho _{i},\pi _{i}\}\right) =\sum_{j=1}^{n}q_{j}\chi
_{\Phi _{j}}\left( \{\rho _{i},\pi _{i}\}\right) .\qquad \square
\end{equation*}

We shall call $\Phi $ the \textit{direct sum mixture} of the channels $
\{\Phi _j\}_{j=1}^n.$

\textbf{Proposition 2.} \textit{Let }$\Psi $\textit{\ be an arbitrary
channel. The inequality (\ref{sub-add}) holds in each of the following
cases: }

$\mathit{(i)}$ $\Phi $\textit{\ is a noiseless channel;}

$\mathit{(ii)}$ $\Phi $\textit{\ is an entanglement breaking channel;}

$\mathit{(iii)}$ $\Phi $\textit{\ is a direct sum mixture of a noiseless
channel and a channel }$\Phi _0$\textit{\ such that (\ref{sub-add}) holds
for }$\Phi _0$\textit{\ and }$\Psi $\textit{\ (in particular, an
entanglement breaking channel).}

An obvious example of a channel of the type $\mathit{(iii)}$ is erasure
channel.

\textit{Proof.} $\mathit{(i)}$ The proof is a modification of the proof in
\cite{H-QI} of the ''unconstrained'' additivity for two channels with one of
them noiseless, based on the Groenevold-Lindblad-Ozawa inequality \cite{O}
\begin{equation}
H(\sigma )\geq \sum_jp_jH(\sigma _j),  \label{GLO}
\end{equation}
where $\sigma $ is a state of a quantum system before von Neumann
measurement, $\sigma _j$ --- the posterior state with the outcome $j$ and $
p_j$ is the probability of this outcome.

Let $\Phi =\mathrm{Id}$ be the noiseless channel and let $\rho $ be an
arbitrary state in $\mathfrak{S}(\mathcal{H})$. We want to prove that
\begin{equation}
\bar{C}(\mathrm{Id}\otimes \Psi ,\{\rho \}\otimes \{\omega \})=\bar{C}(
\mathrm{Id},\{\rho \})+\bar{C}(\Psi ,\{\omega \})=H(\rho )+\chi _\Psi
(\omega )  \label{s-c-add}
\end{equation}
Let $\{\mu _i,\sigma _i\}$ be an ensemble of states in $\mathfrak{S}(
\mathcal{H\otimes K})$ with $\sum_i\mu _i\sigma _i^\Phi =\rho ,\sum_i\mu
_i\sigma _i^\Psi =\omega $. By subadditivity of quantum entropy
\begin{equation}
\begin{array}{c}
\chi _{\mathrm{Id}\otimes \Psi }(\{\mu _i,\sigma _i\})=H(\mathrm{Id}\otimes
\Psi (\sum\limits_i\mu _i\sigma _i))-\sum\limits_i\mu _iH(\mathrm{Id}\otimes
\Psi (\sigma _i)) \\
\\
\leq H(\rho )+H(\Psi (\omega ))-\sum\limits_i\mu _iH(\mathrm{Id}\otimes \Psi
(\sigma _i)).
\end{array}
\label{a}
\end{equation}
Consider the measurement, defined by the observable $\{|e_j\rangle \langle
e_j|\otimes I_{\mathcal{K}}\}$, where $\{|e_j\rangle \}$ is an orthonormal
basis in $\mathcal{H}$. By (\ref{GLO}) we obtain
\begin{equation*}
H(\mathrm{Id}\otimes \Psi (\sigma _i))\geq \sum_jp_{ij}H(\Psi (\sigma
_{ij}^\Psi )),\quad \text{for all }i,
\end{equation*}
where $p_{ij}=\langle e_j|\sigma _i|e_j\rangle $ and $\sigma
_{ij}=p_{ij}^{-1}|e_j\rangle \langle e_j|\otimes I_{\mathcal{K}}\cdot \sigma
_i\cdot |e_j\rangle \langle e_j|\otimes I_{\mathcal{K}}$. Note that $
\sum_jp_{ij}\sigma _{ij}^\Psi =\sigma _i^\Psi $ and $\sum_{ij}\mu
_ip_{ij}\sigma _{ij}^\Psi =\omega $. This and previous inequality show that
two last terms in (\ref{a}) do not exceed $\chi _\Psi (\{\mu _ip_{ij},\sigma
_{ij}^\Psi \})$ and, hence, $\chi _\Psi (\omega )$. With this observation (
\ref{a}) implies (\ref{s-c-add}) and hence the proof is complete.

$\mathit{(ii)}$ See \cite{Sh-e-b-c} where the additivity conjecture for two
unconstrained channels with one of them is entanglement breaking was proved.
In the proof of this theorem the subadditivity property of the $\chi$
-function was in fact established. We can also deduce the subadditivity of
the $\chi $-function from the unconstrained additivity with the help of
corollary 2 (see Sec. 5 below). One should only verify that entanglement
breaking property of a channel implies similar property of Shor's extension
for that channel.

$\mathit{(iii)}$ Let $\Phi _{q}=q\mathrm{Id}\oplus (1-q)\Phi _{0}$. For an
arbitrary channel $\Psi $ we have $\Phi _{q}\otimes \Psi =q(\mathrm{Id}
\otimes \Psi )\oplus (1-q)(\Phi _{0}\otimes \Psi )$. By using lemma 3 and
subadditivity of the functions $\chi _{\mathrm{Id}\otimes \Psi }$ and $\chi
_{\Phi _{0}\otimes \Psi }$,
\begin{equation*}
\begin{array}{c}
\chi _{\Phi _{q}\otimes \Psi }(\sigma )\leq q\chi _{\mathrm{Id}\otimes \Psi
}(\sigma )+(1-q)\chi _{\Phi _{0}\otimes \Psi }(\sigma ) \\
\\
\leq q\chi _{\mathrm{Id}}(\sigma ^{\Phi })+q\chi _{\Psi }(\sigma ^{\Psi
})+(1-q)\chi _{\Phi _{0}}(\sigma ^{\Phi })+(1-q)\chi _{\Psi }(\sigma ^{\Psi
}) \\
\\
=qH(\sigma ^{\Phi })+(1-q)\chi _{\Phi _{0}}(\sigma ^{\Phi })+\chi _{\Psi
}(\sigma ^{\Psi })=\chi _{\Phi _{q}}(\sigma ^{\Phi })+\chi _{\Psi }(\sigma
^{\Psi }),
\end{array}
\end{equation*}
where the last equality follows from the existence of a \textit{pure} state
ensemble on which the maximum in the definition of $\chi _{\Phi _{0}}(\sigma
^{\Phi })$ is achieved. $\square $

\section{Shor's channel extension}

Let $\Phi $ be a channel from $\mathfrak{S}(\mathcal{H})$ to $\mathfrak{S}(
\mathcal{H}^{\prime})$, and let $E$ be an operator in $\mathfrak{B}(\mathcal{
\ H}),0\leq E\leq I$. Let $q\in [0;1]$ and $d\in \mathbb{N} =\{1,2,\dots \}.$
Shor's channel extension $\widehat{\Phi }$ with probability $1-q$ acts as
the channel $\Phi $ and with probability $q$ makes a measurement in $
\mathcal{H}$ with the outcomes $\left\{ 0,1\right\} $ corresponding to the
resolution of the identity $\left\{ E^{\bot },E\right\} ,$ where we denote $
E^{\bot }=I-E.$ If the outcome is $1,$then $\log d$ classical bits are sent
to the receiver, otherwise -- a failure signal \cite{Sh-e-a-q}. Later $q$
will tend to zero while $d$ -- to infinity, such that $q\log d=\lambda $
will be constant. The channel $\widehat{\Phi }$ will then mostly act on
input states $\rho $ as $\Phi ,$ at the same time rarely sending a lot of
classical information at the rate proportional to the value $\mathrm{Tr}\rho
E,$ which to some extent explains its relation to the capacity of channel $
\Phi $ with constrained inputs to be explored in this section.

Translating the definition into algebraic language, consider the following
channel $\widehat{\Phi }(E,q,d)$, which maps states on $\mathfrak{B}(
\mathcal{H})\otimes \mathbf{C}^d$ into states on $\mathfrak{B}(\mathcal{H}
^{\prime })\oplus \mathbf{C}^{d+1}$, where $\mathbf{C}^d$ is the commutative
algebra of complex $d$-dimensional vectors describing a classical system. By
using the isomorphism of $\mathfrak{B}(\mathcal{H})\otimes \mathbf{C}^d$
with the direct sum of $d$ copies of $\mathfrak{B}(\mathcal{H})$, any state
in $\mathfrak{B}(\mathcal{H})\otimes \mathbf{C}^d$ can be represented as an
array $\{\rho _j\}_{j=1}^d$ of positive operators in $\mathfrak{B}(\mathcal{
\ H })$ such that $\mathrm{Tr}\sum_{j=1}^d\rho _j=1$. The action of the
channel $\widehat{\Phi }(E,q,d)$ on the state $\widehat{\rho }=\{\rho
_j\}_{j=1}^d$ with $\rho =\sum_{j=1}^d\rho _j$ is defined by
\begin{equation*}
\widehat{\Phi }(E,q,d)(\widehat{\rho })=(1-q)\Phi _0(\widehat{\rho })\oplus
q\Phi _1(\widehat{\rho }),
\end{equation*}
where $\Phi _0(\widehat{\rho })\!=\Phi (\rho )\in \mathfrak{S}(\mathcal{H}
^{\prime })$ and $\Phi _1(\widehat{\rho })=[\mathrm{Tr}\rho E^{\bot },\,
\mathrm{Tr}\rho_{1}E,...,\mathrm{Tr}\rho_{d}E]\in \mathbf{C}^{d+1}\!.$ Note
that $\Phi _0$ and $\Phi _1$ are channels from $\mathfrak{B}(\mathcal{H}
)\otimes \mathbf{C}^d$ to $\mathfrak{B}(\mathcal{H}^{\prime })$ and to $
\mathbf{C}^{d+1}$ correspondingly. The input state space of the channel $
\widehat{\Phi }(E,q,d)$ will be denoted $\mathfrak{S}_{\widehat{\Phi }}$.

\textbf{Remark 2.} More precisely, since in this paper channel means a map
defined on the algebra of all operators in the input Hilbert space, the
action of $\widehat{\Phi }(E,q,d)$ should be extended correspondingly. Then $
\mathbf{C}^d$ is considered as the algebra of diagonal matrices acting in $
d- $dimensional Hilbert space $\mathcal{H}_d,$ and the input algebra of the
channel $\mathfrak{B}(\mathcal{H})\otimes \mathbf{C}^d\subset \mathfrak{B}(
\mathcal{H}\otimes \mathcal{H}_{d}),$ while the output algebra $\mathfrak{B}
(\mathcal{H}^{\prime })\oplus \mathbf{C}^{d+1}\subset \mathfrak{B}(\mathcal{
H }^{\prime }\oplus \mathcal{H}_{d+1}).$ The action of $\widehat{\Phi }
(E,q,d) $ can then be naturally extended to the whole of $\mathfrak{B}(
\mathcal{H} \otimes \mathcal{H}_{d})$ by letting $\widehat{\Phi } $ vanish
on the elements $A\otimes B,$ where $A\in \mathfrak{B}(\mathcal{H})$ and $B$
is any matrix with zeroes on the diagonal, acting in $\mathcal{H}_d.$ This
is described in \cite{Sh-e-a-q} by saying that the first action of $\widehat{
\Phi }(E,q,d)$ is to make a measurement in the canonical basis of $\mathcal{
H }_d$.

\textbf{Proposition 3.}\textit{\ Let} $\Psi :\mathfrak{S}(\mathcal{K}
)\mapsto \mathfrak{S}(\mathcal{K}^{\prime })$ \textit{be an arbitrary } $
\mathcal{B}$\textit{-constrained channel. Consider the channel }$\widehat{
\Phi }(E,q,d)\otimes \Psi $\textit{. Then}
\begin{equation*}
\begin{array}{c}
\!\!\!\left| \bar{C}\!\left( \widehat{\Phi }(E,q,d)\otimes \Psi ,\mathfrak{S}
_{\widehat{\Phi }}\otimes \mathcal{B}\right) -\!\!\!\!\max\limits_{\sigma :
\mathrm{Tr}_{\mathcal{H}}\sigma \in \mathcal{B}}\left[ (1\!-\!q)\chi _{\Phi
\otimes \Psi }(\sigma )+q\log d\mathrm{Tr}\,\sigma (E\otimes I_{\mathcal{K}
})\right] \right| \\
\\
\leq q(\log \dim \mathcal{\mathcal{K}}^{\prime }+1).
\end{array}
\end{equation*}
\textit{Proof.} Due to the representation
\begin{equation}
\widehat{\Phi }(E,q,d)\otimes \Psi =(1-q)\left( \Phi _0\otimes \Psi \right)
\oplus q\left( \Phi _1\otimes \Psi \right) ,  \label{d-s-rep}
\end{equation}
lemma 3 reduces the calculation of the quantity $\chi _{\widehat{\Phi }
(E,q,d)\otimes \Psi }$ for any ensemble of input states to the calculation
of the quantities $\chi _{\Phi _0\otimes \Psi }$ and $\chi _{\Phi _1\otimes
\Psi }$ for this ensemble.

Note that any state $\widehat{\sigma }$ in $\mathfrak{B}(\mathcal{H})\otimes
\mathbf{C}^d\otimes \mathfrak{B}(\mathcal{K})$ can be represented as an
array $\{\sigma _j\}_{j=1}^d$ of positive operators in $\mathfrak{B}(
\mathcal{H}\otimes \mathcal{K})$ such that $\mathrm{Tr}\sum_{j=1}^d\sigma
_j=1$. Denote by $\delta _j(\sigma )$ the array $\hat{\sigma}$ with the
state $\sigma $ in the $j$-th position and with zeroes in other places.

It is known that for any channel there exists a pure state optimal ensemble
\cite{Sch-West-1} and that the image of the average state of any optimal
ensemble is the same (this follows from corollary 1). These facts and
symmetry arguments imply existence of an optimal ensemble for the channel $
\widehat{\Phi }(E,q,d)\otimes \Psi $ consisting of the states $\widehat{
\sigma }_{i,j}=\delta _j(\sigma _i)$ with the probabilities $\widehat{\mu }
_{i,j}=d^{-1}\mu _i$, where $\{\mu _i,\sigma _i\}$ is an ensemble of states
in $\mathfrak{S}( \mathcal{H\otimes K})$ (cf. \cite{Sh-e-a-q}). Let $
\widehat{\sigma }_{\mathrm{av}}=\sum_{i,j}\widehat{\mu }_{i,j}\widehat{
\sigma }_{i,j}$ and $\sigma _{\mathrm{av}}=\sum_i\mu _i\sigma _i$ be the
averages of these ensembles. Note that $\widehat{\sigma }_{\mathrm{av}
}=[d^{-1}\sigma _{\mathrm{av}},...,d^{-1}\sigma _{\mathrm{av}}]$.

The action of the channel $\Phi _0\otimes \Psi $ on the state $\widehat{
\sigma }=\left[ \sigma _j\right] _{j=1}^d$ with $\sigma =\sum_{i=1}^d\sigma
_i$ is
\begin{equation*}
\Phi _0\otimes \Psi (\widehat{\sigma })=\Phi \otimes \Psi (\sigma ).
\end{equation*}
Hence $\Phi _0\otimes \Psi (\widehat{\sigma }_{i,j})=\Phi \otimes \Psi
(\sigma _i)$ and
\begin{equation}
\chi _{\Phi _0\otimes \Psi }(\{\widehat{\mu }_{i,j},\widehat{\sigma }
_{i,j}\})=\chi _{\Phi \otimes \Psi }(\{\mu _i,\sigma _i\}).  \label{chi-11}
\end{equation}
Let us prove that
\begin{equation}
\begin{array}{c}
\chi _{\Phi _1\otimes \Psi }(\{\widehat{\mu }_{i,j},\widehat{\sigma }
_{i,j}\})=\log d\mathrm{Tr}\sigma _{\mathrm{av}}(E\otimes I_{\mathcal{K}
})+f_\Psi ^E(\{\mu _i,\sigma _i\}),
\end{array}
\label{chi-12}
\end{equation}
where $0\leq f_\Psi ^E(\{\mu _i,\sigma _i\})\leq \log \dim \mathcal{K}
^{\prime }+1$. It is easy to see that the action of the channel $\Phi
_1\otimes \Psi $ on the state $\widehat{\sigma }=\left[ \sigma _j\right]
_{j=1}^d$ with $\sigma =\sum_{i=1}^d\sigma _i$ is
\begin{equation*}
\Phi _1\otimes \Psi (\widehat{\sigma })=[\Psi _{E^{\bot }}(\sigma ),\Psi
_E(\sigma _1),...,\Psi _E(\sigma _d)],
\end{equation*}
where $\Psi _A(\cdot )=\mathrm{Tr}_{\mathcal{H}}(A\otimes I_{\mathcal{K}})(
\mathrm{Id}\otimes \Psi )(\cdot )$ is a completely positive
trace-nonincreasing map from $\mathfrak{B}(\mathcal{H\otimes K})$ into $
\mathfrak{B}(\mathcal{\ K^{\prime }})$, ($A=E,E^{\bot },$ and $\mathrm{Id}$
is the identity map on $\mathfrak{S}( \mathcal{H})$).

Therefore,
\begin{equation}
H(\Phi _1\otimes \Psi (\widehat{\sigma }_{i,j}))=H(\Psi _{E^{\bot }}(\sigma
_i))+H(\Psi _E(\sigma _i)),  \label{H-Phi-12-S-i}
\end{equation}
and
\begin{equation*}
\begin{array}{c}
\Phi _1\otimes \Psi (\widehat{\sigma }_{\mathrm{av}})=\sum\limits_{i,j}
\widehat{\mu }_{i,j}\Phi _1\otimes \Psi (\widehat{\sigma }_{i,j}) \\
=[\Psi _{E^{\bot }}(\sigma _{\mathrm{av}}),d^{-1}\Psi _E(\sigma _{\mathrm{av}
}),...,d^{-1}\Psi _E(\sigma _{\mathrm{av}})],
\end{array}
\end{equation*}
Due to this
\begin{equation}
H(\Phi _1\otimes \Psi (\widehat{\sigma }_{\mathrm{av}}))=\log d\mathrm{\ \
\mathrm{Tr}}\Psi _E(\sigma _{\mathrm{av}})+H(\Psi _E(\sigma _{\mathrm{av}
}))+H(\Psi _{E^{\bot }}(\sigma _{\mathrm{av}})).  \label{H-Phi-12-S}
\end{equation}
Using (\ref{H-Phi-12-S-i}), (\ref{H-Phi-12-S}) and $\mathrm{Tr}\Psi
_E(\sigma )=\mathrm{Tr}\sigma (E\otimes I_{\mathcal{K}})$, we obtain
\begin{equation}
\begin{array}{c}
\chi _{\Phi _1\otimes \Psi }(\{\widehat{\mu }_{i,j},\widehat{\sigma }
_{i,j}\})=\log d\,\mathrm{Tr}\sigma _{\mathrm{av}}(E\otimes I_{\mathcal{K}})
\\
\\
+H(\Psi _E(\sigma _{\mathrm{av}}))+H(\Psi _{E^{\bot }}(\sigma _{\mathrm{av}
}))-\sum\limits_i\mu _i(H(\Psi _E(\sigma _i))+H(\Psi _{E^{\bot }}(\sigma
_i))) \\
\\
=\log d\mathrm{Tr}\sigma _{\mathrm{av}}(E\otimes I_{\mathcal{K}})+\chi
_{\Psi _E}(\{\mu _i,\sigma _i\})+\chi _{\Psi _{E^{\bot }}}(\{\mu _i,\sigma
_i\}).
\end{array}
\label{chi-12+}
\end{equation}
Using the inequalities $0\leq H(S)\leq \mathrm{Tr}S(\log \dim \mathcal{H}
-\log \mathrm{Tr}S)$ for any positive operator $S\in \mathcal{B}(\mathcal{H}
),$ and $h_2(x)=x\log x+(1-x)\log (1-x)\leq 1,$ it is possible to show that
\begin{equation}
f_\Psi ^E(\{\mu _i,\sigma _i\}):=\chi _{\Psi _E}(\{\mu _i,\sigma _i\})+\chi
_{\Psi _{E^{\bot }}}(\{\mu _i,\sigma _i\})\leq \log \dim \mathcal{K}^{\prime
}+1,  \label{est}
\end{equation}
hence we obtain (\ref{chi-12}).

Lemma 3 with (\ref{chi-11}) and (\ref{chi-12}) imply
\begin{equation*}
\begin{array}{c}
\chi _{\widehat{\Phi }(E,q,d)\otimes \Psi }(\{\widehat{\mu }_{i,j},\widehat{
\sigma }_{i,j}\})\!=\!(1-q)\chi _{\Phi _{0}\otimes \Psi }(\{\widehat{\mu }
_{i,j},\widehat{\sigma }_{i,j}\})+q\chi _{\Phi _{1}\otimes \Psi }(\{\widehat{
\mu }_{i,j},\widehat{\sigma }_{i,j}\}) \\
\\
=(1-q)\chi _{\Phi \otimes \Psi }(\{\mu _{i},\sigma _{i}\})+q\log d\mathrm{Tr}
\sigma _{\mathrm{av}}(E\otimes I_{\mathcal{K}})+qf_{\Psi }^{E}(\{\mu
_{i},\sigma _{i}\}).
\end{array}
\end{equation*}
The last equality with (\ref{est}) completes the proof. $\square $

\textbf{Theorem 2.} \textit{Let} $\Phi :\mathfrak{S}(\mathcal{H})\mapsto
\mathfrak{S}(\mathcal{H}^{\prime })$ \textit{and} $\Psi :\mathfrak{S}(
\mathcal{K})\mapsto \mathfrak{S}(\mathcal{K}^{\prime })$ \textit{be
arbitrary channels with the fixed constraint on the second one defined by a
closed set }$\mathcal{B}$\textit{. The following statements are equivalent:}

$\mathit{(i)}$ \textit{The additivity }(\ref{addit})\textit{\ holds for the}
$\mathcal{A}$-\textit{constrained channel }$\Phi $\textit{\ with arbitrary
closed} $\mathcal{A}\in \mathfrak{S}(\mathcal{H})$ \textit{and the} $
\mathcal{B}$-\textit{constrained channel }$\Psi$;

$\mathit{(ii)}$\textit{The additivity holds asymptotically for the sequence
of the channels }$\{\widehat{\Phi }(E,\lambda /\log d,d)\}_{d\in \mathbb{N}
} $ \textit{with arbitrary operator} $0\leq E\leq I$ \textit{and arbitrary
nonnegative number} $\lambda$ \textit{\ (without constraints) and the }$
\mathcal{\ B}$-\textit{constrained channel }$\Psi $, \textit{in the sense
that}
\begin{equation*}
\lim_{d\rightarrow +\infty }\bar{C}(\widehat{\Phi }(E,\lambda /\log
d,d)\otimes \Psi ,\mathfrak{S}_{\widehat{\Phi }}\otimes \mathcal{B}
)=\lim_{d\rightarrow +\infty }\bar{C}(\widehat{\Phi }(E,\lambda /\log d,d))+
\bar{C}(\Psi ,\mathcal{B}).
\end{equation*}

\textit{Proof.} Note, first of all, that for an operator $0\leq E\leq I$ and
a number $\lambda \geq 0$ proposition 3 implies
\begin{equation}
\lim\limits_{d\rightarrow +\infty }\bar{C}(\widehat{\Phi }(E,\lambda /\log
d,d))=\max_\rho \left[ \chi _\Phi (\rho )+\lambda \mathrm{Tr}\rho E\right]
\label{asymp-1}
\end{equation}
and
\begin{equation}
\lim\limits_{d\rightarrow +\infty }\bar{C}(\widehat{\Phi }(E,\lambda /\log
d,d)\otimes \Psi ,\mathfrak{S}_{\widehat{\Phi }}\otimes \mathcal{B}
)=\max_{\sigma :\mathrm{Tr}_{\mathcal{H}}\sigma \in \mathcal{B}}\left[ \chi
_{\Phi \otimes \Psi }(\sigma )+\lambda \mathrm{\ Tr}\,\sigma (E\otimes I_{
\mathcal{K}})\right]  \label{asymp-2}
\end{equation}
correspondingly.

Begin with $({i})\Rightarrow ({ii})$. Let $\sigma _{\ast }$ be a maximum
point in the right side of (\ref{asymp-1}) and $\alpha =\mathrm{Tr}\sigma
_{\ast }(E\otimes I_{\mathcal{K}})$. By the statement $({i})$ the additivity
holds for the channel $\Phi $ with the constraint $\mathrm{Tr}\rho E^{\bot
}\leq 1-\alpha $ and the $\mathcal{B}$ -constrained channel $\Psi $. So
there exist such states $\rho $ and $\omega \in \mathcal{B}$ that $\mathrm{
Tr }\rho E\geq \alpha $ and $\chi _{\Phi }(\rho )+\chi _{\Psi }(\omega )\geq
\chi _{\Phi \otimes \Psi }(\sigma _{\ast })$. Hence
\begin{equation*}
\begin{array}{c}
\max\limits_{\sigma :\,\sigma ^{\Psi }\in \mathcal{B}}\left[ \chi _{\Phi
\otimes \Psi }(\sigma )+\lambda \mathrm{Tr}\sigma (E\otimes I_{\mathcal{K}
})\,\right] =\chi _{\Phi \otimes \Psi }(\sigma _{\ast })+\lambda \mathrm{Tr}
\,\sigma _{\ast }(E\otimes I_{\mathcal{K}}) \\
\leq \chi _{\Phi }(\rho )+\chi _{\Psi }(\omega )+\lambda \mathrm{Tr}\rho
E\leq \max\limits_{\rho }\left[ \chi _{\Phi }(\rho )+\lambda \mathrm{Tr}\rho
E\right] +\bar{C}(\Psi ;\mathcal{B}).
\end{array}
\end{equation*}
Due to (\ref{asymp-1}) and (\ref{asymp-2}) this means that
\begin{equation*}
\lim_{d\rightarrow +\infty }\bar{C}(\widehat{\Phi }(E,\lambda /\log
d,d)\otimes \Psi ,\mathfrak{S}_{\widehat{\Phi }}\otimes \mathcal{B})\leq
\lim_{d\rightarrow +\infty }\bar{C}(\widehat{\Phi }(E,\lambda /\log d,d))+
\bar{C}(\Psi ,\mathcal{B})
\end{equation*}
which implies $({ii})$.

The proof of $({ii})\Rightarrow ({i})$ is based on lemma 2. Let $\mathcal{A}
^{\mathit{l}}$ be a set defined by the inequality $\mathrm{\ Tr}\rho A\leq
\alpha $ with an operator $0\leq A\leq I$ and a positive number $\alpha $
such that there exists a state $\rho ^{\prime }$ with $\mathrm{Tr}\rho
^{\prime }A<\alpha $. Due to lemma 2 it is sufficient to show that
\begin{equation}
\bar{C}\left( \Phi \otimes \Psi ;\mathcal{A}^{\mathit{l}}\otimes \mathcal{B}
\right) \leq \bar{C}(\Phi ;\mathcal{A}^{\mathit{l}})+\bar{C}(\Psi ;\mathcal{B
}),  \label{add-c-m}
\end{equation}
that is, for all ensembles $\{\mu _{i},\sigma _{i}\}$ in $\mathfrak{S}(
\mathcal{H}\otimes \mathcal{K})$ with the average $\sigma _{\mathrm{av}},$
such that $\mathrm{Tr}\sigma _{\mathrm{av}}^{\Phi }A\leq \alpha $, $\sigma _{
\mathrm{av}}^{\Psi }\in \mathcal{B}$,
\begin{equation}
\chi _{\Phi \otimes \Psi }(\sigma _{\mathrm{av}})\leq \bar{C}(\Phi ;\mathcal{
A}^{\mathit{l}})+\bar{C}(\Psi ;\mathcal{B}).  \label{c-c-add-c}
\end{equation}

Let $\rho _{\mathrm{av}}$ be the average state of the optimal ensemble for
the $\mathcal{A}^{\mathit{l}}$-constrained channel $\Phi $ so that $\bar{C}
(\Phi ;\mathcal{A}^{\mathit{l}})=\chi _{\Phi }(\rho _{\mathrm{av}}).$ Note
that the state $\rho _{\mathrm{av}}$ is the point of maximum of the concave
function $\chi _{\Phi }(\rho )$ with the constraint $\mathrm{Tr}\rho A\leq
\alpha $. By the Kuhn-Tucker theorem (we use the strong version of this
theorem with the Slater condition, which follows from the existence of a
state $\rho ^{\prime }$ such that $\mathrm{Tr}\rho ^{\prime }A<\alpha $ )
\cite{JT}, there exists a nonnegative number $\lambda $, such that $\rho _{
\mathrm{av}}$ is the point of the global maximum of the function $\chi
_{\Phi }(\rho )-\lambda \mathrm{Tr}\rho A$ and the following condition holds
\begin{equation}
\lambda (\mathrm{Tr}A\rho _{\mathrm{av}}-\alpha )=0.  \label{c-a-f}
\end{equation}
It is clear that $\rho _{\mathrm{av}}$ is also the point of the global
maximum of the concave function $\chi _{\Phi }(\rho )+\lambda \mathrm{Tr}
\rho E$, where $E=I-A,$ so that
\begin{equation}
\chi _{\Phi }(\rho )+\lambda \mathrm{Tr}\rho E\leq \chi _{\Phi }(\rho _{
\mathrm{av}})+\lambda \mathrm{Tr}\rho _{\mathrm{av}}E,\quad \forall \rho \in
\mathfrak{S}(\mathcal{H}).  \label{chi-ine}
\end{equation}
Consider the sequence $\widehat{\Phi }(E,\lambda /\log d,d)$. Assumed
asymptotic additivity together with (\ref{asymp-1}) and (\ref{asymp-2})
implies
\begin{equation}
\max_{\sigma }\left[ \chi _{\Phi \otimes \Psi }(\sigma )+\lambda \mathrm{Tr}
\,\sigma (E\otimes I_{\mathcal{K}})\right] =\max_{\rho }\left[ \chi _{\Phi
}(\rho )+\lambda \mathrm{Tr}\rho E\right] +\bar{C}(\Psi ;\mathcal{B}).
\label{asymp-e}
\end{equation}
Due to (\ref{c-a-f}) and (\ref{chi-ine}) we have
\begin{equation}
\max\limits_{\rho }\left[ \chi _{\Phi }(\rho )+\lambda \mathrm{Tr}\rho E
\right] =\chi _{\Phi }(\rho _{\mathrm{av}})+\lambda \mathrm{Tr}\rho _{
\mathrm{av}}(I-A)=\bar{C}(\Phi ;\mathcal{A}^{l})+\lambda (1-\alpha ).
\label{asymp-e-1}
\end{equation}
Hence
\begin{equation*}
\chi _{\Phi \otimes \Psi }(\sigma _{\mathrm{av}})+\lambda \mathrm{Tr}\sigma
_{\mathrm{av}}(E\otimes I_{\mathcal{K}})\,\leq \bar{C}(\Phi ;\mathcal{A}
^{l})+\bar{C}(\Psi ;\mathcal{B})+\lambda (1-\alpha ).
\end{equation*}
Noting that
\begin{equation*}
\mathrm{Tr}\,\sigma _{\mathrm{av}}(E\otimes I_{\mathcal{K}})=\mathrm{Tr}
\sigma _{\mathrm{av}}^{\Phi }(I-A)\geq 1-\alpha ,
\end{equation*}
we obtain (\ref{c-c-add-c}), and hence $({ii})\Rightarrow ({i})$. $\square $

\textbf{Corollary 2.} \textit{The additivity of }$\chi $\textit{-capacity
for the Shor's channel extensions }$\widehat{\Phi }(E,q,d)$\textit{\ and }$
\widehat{\Psi }(F,r,e)$\textit{\ with arbitrary pairs }$(E,q,d)$\textit{\
and }$(F,r,e)$ \textit{\ implies its additivity for the} $\mathcal{A}$
\textit{-constrained channel }$\Phi $\textit{\ and the} $\mathcal{B}$\textit{
\ \ -constrained channel }$\Psi $\textit{\ with arbitrary} $\mathcal{A}
\subset \mathfrak{S}(\mathcal{H})$ \textit{and} $\mathcal{B}\subset
\mathfrak{S}( \mathcal{K})$.

\textit{Proof. }This is obtained by double application of theorem 2. $
\square $

\textbf{Corollary 3.} \textit{If the additivity holds for any two
unconstrained channels then it holds for any two channels with arbitrary
constraints.}

\textbf{Remark 3.} The statement of the corollary 3 could be also deduced by
combining results of \cite{Sh-e-a-q} and \cite{MSW}, but we gave a direct
proof here.

\section{Additive constraints}

Let $A$ be a positive operator in $\mathcal{H}$, and let
\begin{equation*}
A^{(n)}=A\otimes \dots \otimes I_{\mathcal{H}}+\dots +I_{\mathcal{H}}\otimes
\dots \otimes A
\end{equation*}
be the corresponding operator in $\mathcal{H}^{\otimes n}.$ The classical
capacity of the channel $\Phi $ with inputs subject to the additive
constraint
\begin{equation*}
\mathrm{Tr}\rho ^{(n)}A^{(n)}\leq n\alpha ;\quad n=1,2,\dots
\end{equation*}
is shown \cite{H-c-w-c} to be equal to
\begin{equation*}
C(\Phi ;A,\alpha )=\lim_{n\rightarrow \infty }\bar{C}(\Phi ^{\otimes
n};A^{(n)},n\alpha )/n.
\end{equation*}

In \cite{MSW} the following \textit{weak} additivity property was
considered:
\begin{equation}
\bar{C}(\Phi \otimes \Psi ;A\otimes I_{\mathcal{K}}+I_{\mathcal{H}}\otimes
B,\;\gamma )=\max\limits_{\alpha +\beta =\gamma }\left[ \bar{C}(\Phi
;A,\,\alpha )+\bar{C}(\Psi ;B,\,\beta )\right] ,  \label{w-a-p}
\end{equation}
where $\Phi $ and $\Psi $ are channels with the input spaces $\mathcal{H}$
and $\mathcal{K}$, and the corresponding linear constraints $\mathrm{Tr}\rho
A\leq \alpha $ and $\mathrm{Tr}\rho B\leq \beta $. It is easy to see that
the additivity for the two constrained channels in the sense (\ref{addit})
implies the weak additivity (\ref{w-a-p}). The extension of the latter to $n$
channels implies
\begin{equation*}
\bar{C}(\Phi ^{\otimes n};A^{(n)},n\alpha )=n\bar{C}(\Phi ;A,\alpha )
\end{equation*}
and hence the equality $C(\Phi ;A,\alpha )=\bar{C}(\Phi ;A,\alpha ).$
Indeed, the function $f(\alpha )=\bar{C}(\Phi ;A,\alpha )$ defined by (\ref
{ccap}) is nondecreasing and concave (see Appendix, II), whence
\begin{equation*}
\max_{\alpha _{1}+\dots +\alpha _{n}=n\alpha }\left[ f(\alpha _{1})+\dots
+f(\alpha _{n})\right]
\end{equation*}
is achieved for $\alpha _{1}=\dots =\alpha _{n}=\alpha .$

The weak additivity conjecture for constrained channels becomes equivalent
to the additivity conjecture in the sense of this paper when this weak
additivity holds true for \textit{any} two channels. Indeed, the latter
implies global additivity for channels without constraints, from which
global additivity for constrained channels follows by corollary 3.

Needless to say, however, that in applications constraints usually arise
when the channel space is infinite-dimensional and the constraint operators
are unbounded. The finite dimensionality (implying boundedness of the
constraint operators) is crucial in this paper, and relaxing this
restriction is both interesting and nontrivial problem.

\section{Appendix}

I. The main property underlying the proof of the lemma 1 is the concavity of
the function $\chi _{\Phi }(\rho )$ on $\mathfrak{S}(\mathcal{H})$. This
function may not be smooth, therefore we will use non-smooth convex analysis
arguments instead of derivatives calculations.

Consider the Banach space $\mathfrak{B}_{h}(\mathcal{H})$ of all Hermitian
operators on $\mathcal{H}$ and the concave extension $\widehat{\chi }_{\Phi
} $ of the function $\chi _{\Phi }$ to $\mathfrak{B}_{h}(\mathcal{H})$,
defined by:
\begin{equation*}
\widehat{\chi _{\Phi }}(\rho )=\left\{
\begin{array}{cc}
\lbrack \mathrm{Tr}\rho ]\cdot \chi _{\Phi }([\mathrm{Tr}\rho ]^{-1}\rho ),
& \rho \in \mathfrak{B}_{+}(\mathcal{H}); \\
-\infty , & \rho \in \mathfrak{B}_{h}(\mathcal{H})\backslash \mathfrak{B}
_{+}(\mathcal{H}),
\end{array}
\right.
\end{equation*}
where $\mathfrak{B}_{+}(\mathcal{H})$ is the convex cone of positive
operators in $\mathcal{H}$. The function $\widehat{\chi }_{\Phi }$ is
bounded in a neighborhood of any internal point of $\mathfrak{B}_{+}(
\mathcal{H})$ (and, hence, by the concavity it is continuous at all internal
points of $\mathfrak{B}_{+}(\mathcal{H})$, which are nondegenerate positive
operators, see \cite{JT}, 3.2.3).

By the assumption $\rho _{0}$ is an internal point of the cone $\mathfrak{B}
_{+}(\mathcal{H})$. Hence, the convex function $-\widehat{\chi }_{\Phi }$ is
continuous at $\rho _{0}$. Due to the continuity, the subdifferential of the
convex function $-\widehat{\chi }_{\Phi }$ at the point $\rho _{0}$ is not
empty (see \cite{JT}, 4.2.1). This means that there exists a linear function
$l(\rho )$ such that $\rho _{0}$ is the minimum point of the function $-
\widehat{\chi }_{\Phi }(\rho )-l(\rho )$. Any linear function on $\mathfrak{
\ \ \ \ B}_{h}(\mathcal{H})$ has the form $l(\rho )=\mathrm{Tr}A\rho $ for
some $A\in \mathfrak{B}_{h}(\mathcal{H})$. Hence, $\rho _{0}$ is also the
minimum point of the function $-\widehat{\chi }_{\Phi }(\rho )\;$ under the
conditions $\mathrm{Tr}A\rho =\alpha =\mathrm{Tr}A\rho _{0}$ and $\mathrm{Tr}
\rho =1$. Introduce the operator $A^{\prime }=\frac{1}{2}[\Vert A\Vert
^{-1}A+I]$ and the number $\alpha ^{\prime }=\frac{1}{2}[\Vert A\Vert
^{-1}\alpha +1]$. The linear variety defined by the conditions $\mathrm{Tr}
\rho A=\alpha $ and $\mathrm{Tr}\rho =1$ coincides with that defined by the
conditions $\mathrm{Tr}A^{\prime }\rho =\alpha ^{\prime }$ and $\mathrm{Tr}
\rho =1$. Therefore, $\rho _{0}$ is the minimum point of the function $-
\widehat{\chi }_{\Phi }(\rho )$ under the conditions $\mathrm{Tr}A^{\prime
}\rho =\alpha ^{\prime }$ and $\mathrm{Tr}\rho =1$, and, hence, $\rho _{0}$
is the maximum point of the function $\chi _{\Phi }(\rho )$ under the
condition $\mathrm{Tr}A^{\prime }\rho =\alpha ^{\prime }$. By concavity of
the function $\chi _{\Phi }(\rho )$ it implies that $\rho _{0}$ is the
maximum point of the function $\chi _{\Phi }(\rho )$ under the condition
either $\mathrm{Tr}A^{\prime }\rho \leq \alpha^{\prime}$ or $\mathrm{Tr}
A^{\prime }\rho \geq \alpha^{\prime}$ (see n. II below). By noting that $
0\leq A^{\prime }\leq I$ and setting $A$ and $\alpha $ to be equal to $
A^{\prime }$ and $\alpha ^{\prime }$ in the first case and to $I-A^{\prime }$
and $1-\alpha ^{\prime } $ in the second, we complete the proof of the lemma
1. \vskip10pt II. If $F(x)$ is a concave continuous function and $l(x)$ is a
linear function on a compact convex subset of a finite dimensional vector
space, then the function
\begin{equation*}
f(\alpha )=\max_{x:l(x)=\alpha }F(x)
\end{equation*}
is concave. Indeed, assume $f(\alpha )$ is not, then there exist $\alpha
_{1},\alpha _{2}$ such that $f(\frac{\alpha _{1}+\alpha _{2}}{2})<$ $\frac{1
}{2}\left[ f(\alpha _{1})+f(\alpha _{2})\right] .$ Let $x_{i}$ be points at
which the maxima are achieved, i. e. $l(x_{i})=\alpha _{i}$ and $f(\alpha
_{i})=F(x_{i}),$ then $l(\frac{x_{1}+x_{2}}{2})=\frac{\alpha _{1}+\alpha
_{2} }{2}$ and $F(\frac{x_{1}+x_{2}}{2})\leq f(\frac{\alpha _{1}+\alpha _{2}
}{2})< $ $\frac{1}{2}\left[ F(x_{1})+F(x_{2})\right] ,$ which contradicts to
the concavity of $F.$ Similar argument applies to the functions $
f_{+}(\alpha )=\max_{x:l(x)\leq \alpha }F(x)$ and $f_{-}(\alpha
)=\max_{x:l(x)\geq \alpha }F(x)$ which are thus also concave$.$

With the same definitions one has either $f(\alpha )=f_{+}(\alpha )$ or $
f(\alpha )=f_{-}(\alpha ),$ for otherwise there exist $x_{1},x_{2}$ such
that
\begin{equation*}
l(x_{1})<\alpha ;\quad F(x_{1})>f(\alpha );\quad l(x_{2})>\alpha ;\quad
F(x_{2})>f(\alpha ).
\end{equation*}
Then taking $\lambda =\frac{l(x_{2})-\alpha }{l(x_{2})-l(x_{1})}$ one has $
0<\lambda <1$, $l(\lambda x_{1}+(1-\lambda )x_{2})=\alpha $ and
\begin{equation*}
F(\lambda x_{1}+(1-\lambda )x_{2})\leq f(\alpha )<\lambda
F(x_{1})+(1-\lambda )F(x_{2}),
\end{equation*}
contradicting the concavity of $F.$

\vskip10pt

{\small {\textit{Acknowledgments.} A.H. thanks P.W. Shor for sending the
draft of his paper \cite{Sh-e-a-q} and acknowledges support from the
Research Program at ZiF, University of Bielefeld, under the supervision of
Prof. R. Ahlswede, where part of this work was done. The authors are
grateful to G. G. Amosov for useful discussion. This work was partially
supported by INTAS grant 00-738.} }

\end{document}